\documentclass[pre,amsmath,floatfix,onecolumn,showpacs]{revtex4}
\usepackage{graphicx}
\usepackage{epsfig}
\usepackage{times}

\newcommand{\eff}{\mathrm{eff}}
\newcommand{\free}{\mathrm{free}}
\newcommand{\sat}{\mathrm{sat}}
\newcommand{\kres}{\kappa_{\mathrm{res}}}
\newcommand{\kpb}{\kappa_{\mathrm{PB}}}

\newcommand{\Rws}{R}

\newcommand{\VECr}{{\mathbf{r}}}
\newcommand{\be}{\begin{equation}}
\newcommand{\ee}{\end{equation}}

\begin{document}

\title{Alexander's prescription for colloidal charge renormalization}

\author{E. Trizac} \affiliation{Laboratoire de Physique Th{\'e}orique,
  UMR 8627, B{\^a}timent 210, Universit{\'e} Paris-Sud, 91405 Orsay
  Cedex, France} 
\author{L. Bocquet} \affiliation{DPM, Universit\'e Claude Bernard Lyon 1, 43
  bld du 11 novembre 1918, 69622 Villeurbanne Cedex, France }
\author{M. Aubouy} \affiliation{SI3M., DRFMC, DSM/CEA-Grenoble, UMR
  5819 (CNRS, UJF), 17 rue des Martyrs, 38054 Grenoble Cedex 9, France }
\author{H.H. von Gr\"unberg} \affiliation{Fachbereich
  Physik, Universit\"at Konstanz, 78457 Konstanz, Germany } 
\date{\today}
\begin{abstract}
  The interactions between charged colloidal particles in an
  electrolyte may be described by usual Debye-H\"uckel theory provided
  the source of the electric field is suitably renormalized. For
  spherical colloids, we reconsider and simplify the treatment of the
  popular proposal put forward by Alexander {\it et al.} [J. Chem.
  Phys. {\bf 80}, 5776 (1984)], which has proven efficient in
  predicting renormalized quantities (charge and salt content).  We
  give explicit formulae for the effective charge and describe the
  most efficient way to apply Alexander's et al. renormalization
  prescription in practice. Particular attention is paid to the
  definition of the relevant screening length, an issue that appears
  confuse in the literature.
\end{abstract}
\maketitle
\section{Introduction}

The concept of charge renormalization is widely used in theories on
effective interactions in charge-stabilized colloidal suspensions
\cite{reviewBelloni97,levin:rev,Likos,Quesada}. The basic idea 
is to consider the highly charged colloidal particle plus
parts of the surrounding cloud of oppositely charged microions as an
entity, carrying a renormalized charge $Z_{\eff}$ which may be orders of
magnitude smaller than the actual bare charge $Z$ of the colloidal
particle.  Replacing $Z$ by $Z_{\eff}$, one can then use simple
Debye-H\"uckel-like theories to calculate the effective interaction
between two such colloidal particles in suspension. The pioneering
work on colloidal charge renormalization is the paper by Alexander et
al. \cite{alexander} who proposed to calculate effective charges by
finding, within the framework of the Poisson-Boltzmann cell model, the
optimal linearized electrostatic potential matching the non-linear one
at the cell boundary. In \cite{alexander}, the prescription how to achieve
this task, is given essentially in form of a numerical recipe which is
often not the most convenient way possible. Making use of simple
analytical expressions for effective charges recently suggested
\cite{trizac1,trizac2}, we here describe an efficient way to
realize Alexander's et al. prescription for charge renormalization.
In the infinite dilution limit of a colloid immersed in an infinite
sea of electrolyte, explicit and accurate analytical expressions 
have been obtained only recently \cite{JPA},
when the Debye length is smaller than the colloid size (spherical 
or cylindrical).

In order to illustrate the general significance of this celebrated
charge renormalization concept, but also to demonstrate how broad its
potential field of application is, we have selected a few recent
studies which are -- in one way or the other -- concerned with
Alexander's et al. prescription. Experimentally, a variety of techniques can
be used to determine effective charges of surfaces and spherical
colloidal particles \cite{Behrens,Wette,vonGru3}, electrophoresis
measurements \cite{Garbow,Wette2} being certainly one of the most
important. Other experiments, focusing on the colloid aggregation
behavior \cite{Fernand,Groenew}, give direct evidence of the important
role of counterion condensation and charge renormalization in
colloidal systems. On the theoretical side, Yukawa-like interaction
potentials with effective charges from Alexander's et al. prescription are
often used as the standard reference curves in calculations of
effective interactions between charged colloids in suspensions
\cite{Lobaski,Terao2,Anta,Mukherj,Ulander}, where, most recently, even
discrete solvent effects \cite{Allahya,Allahya2,Burak} and density
fluctuation effects \cite{Lukatsk} are considered. Charge
renormalization and nonlinear screening effects are also important
when it comes to investigating the phase-behavior of charge-stabilized
colloidal suspensions, be it the yet unsettled question of a possible
gas-liquid phase-coexistence
\cite{Diehl,Schmitz1,vonGru,deserno,Tamashiro,Tellez}, or the
solid-liquid phase-behavior \cite{Liu,Schmitz}. 3D charged colloidal
crystals are rheo-optically investigated in \cite{Okubo}, while defect
dynamics and light-induced melting in 2D colloidal systems is in the
focus of the studies in \cite{Pertsin} and \cite{Beching2,Beching},
respectively.  Theories dealing with dynamic properties of charged
colloidal suspensions are presented, for example, in \cite{Nagele1}.
Another field of application are complexation problems: the
complexation of a polyelectrolyte/DNA with spherical macroions
\cite{Nguyen3,Nguyen}, or, vice versa, the complexation of macroions
with oppositely charged polyelectrolytes
\cite{Nguyen2,Schiess,Zhulina}. Finally, we mention usage of
Alexander's et al. work in recent papers on microgels \cite{Levin}, on the
Rayleigh instability of charged droplets \cite{DDeserno}, on charged
spherical microemulsion and micellar systems \cite{Evilevi,Piazza},
and, as a last and more exotic example, in the theory of dusty plasmas
\cite{Zagorod}. We also note that the phenomenological 
concept of a Stern layer \cite{Stern} --successfully applied
in many studies on adsorption and micellization  of ionic
surfactants \cite{Kalinin}-- is reminiscent of the renormalization
procedure under scrutiny here.

In section \ref{sec:cell}, the general framework is presented, while
Alexander's et al. prescription is revisited in section \ref{sec:alex}.
Within such an approach, a natural screening parameter $\kpb$ appears,
that plays the role of an inverse Debye length. In the literature, by
analogy with the behavior of a bulk system, this parameter is often
related to the mean (effective) micro-ion density $n^*$ in the
suspension through the familiar expression $\kpb^2 = 8 \pi \lambda_B
n^*$, where $\lambda_B$ is the Bjerrum length.  Such a relation
follows from electroneutrality in the latter situation, but becomes
incorrect in the confined geometry we shall be interested in.
Particular emphasis will be put on this issue, when the colloidal
suspension is dialyzed against a salt reservoir (section
\ref{ssec:salt}), or put in the opposite limit of complete
de-ionization (section \ref{ssec:nosalt}).

\section{Cell-model, Poisson-Boltzmann theory and Alexander's et al. prescription}
\label{sec:cell}

We consider a fluid of highly charged colloidal spheres having a
radius $a$ and carrying each a total charge $-Ze$ ($e$ is the
elementary charge). These colloids are suspended in a structureless
medium of relative (CGS) dielectric constant $\varepsilon$ and
temperature $T$ ($\beta = 1/kT$), characterized by the Bjerrum
length, $\lambda_B = \beta e^{2}/\varepsilon$ . The suspension is
dialyzed against a reservoir of monovalent salt ions of a given pair
concentration $c_{s}$ (the counterions are also assumed
monovalent).  
Due to the fact that the colloid compartment
is already occupied by the counterions originating from the macroions,
the salt ion concentration in the reservoir, $c_s$, is always higher than the
average salt ion density $n_{s}$ in the colloid compartment, an effect
which is known as the Donnan-effect \cite{Dubois,trizac40,deserno}.
Note that we not only have $n_s< c_s$ but also $n_s/(1-\eta) < c_s$,
where $\eta$ is the volume fraction of the colloids [the mean density
of salt ions in the volume accessible to microions is thus
$n_s/(1-\eta)$ and not $n_s$].
By contrast, the total concentration of microions in the
system (i.e.  co-ions plus counter-ions) is larger than the total
ion-concentration in the reservoir.

The work of Alexander {\it et al. } 
is based on the Poisson-Boltzmann cell model, an
approximation which attempts to reduce the complicated many particle
problem of interacting charged colloids and microions to an effective
one-colloid-problem \cite{Marcus}.  It rests on the observation that
at not too low volume fractions the colloids -- due to mutual
repulsion -- arrange their positions such that each colloid has a
region around it which is void from other colloids and which looks
rather similar for different colloids.  In other words, the
Wigner-Seitz (WS) cells around two colloids are comparable in shape
and volume.  One now assumes that the total charge within each cell is
exactly zero, that all cells have the same shape, and that one may
approximate this shape such that it matches the symmetry of the
colloid, i.e., spherical cells around spherical colloids.  The cell
radius $\Rws$ is chosen such that $\eta=(a/\Rws)^3$ equals the volume
(or packing) fraction occupied by the colloids.  If one neglects
interactions between different cells, the thermodynamic
potential of the whole suspension is equal to the number of cells
times the thermodynamic potential of one cell.  For a recent and more
detailed description of the cell-model approximation see
Ref.~\cite{DeHo01}.

In a mean-field Poisson-Boltzmann (PB) approach, the key quantity to
calculate is the local electrostatic mean-field potential
$\phi(\VECr)$ (made dimensionless here by multiplication with $\beta
e$), which thanks to the cell-model approximation has to be calculated
in one WS cell only.  It is generated by both the fixed colloidal
charge density as well as the distributions
$n_\pm(\VECr)=c_{s}e^{\mp\phi(\VECr)}$ of mobile monovalent ions, and
follows from the PB equation, which together with the boundary
conditions takes the form,
\begin{equation}
\label{eq:1}
\begin{array}{rcl@{\hspace{1cm}}l}
\nabla^{2} \phi(r) & = & \kres^{2} \sinh \phi(r) & a< r <\Rws \\
\vec{n} \cdot \nabla \phi(r) & = & Z \lambda_{B}/a^{2} & r =a \\
\vec{n} \cdot \nabla \phi(r) & = & 0 & r = \Rws
\end{array} 
\end{equation}
with $\vec{n}$ the outward pointing surface normal and the inverse
screening length $\kres$ defined in terms of the ionic strength of the
reservoir: $\kres^{2} = 8 \pi \lambda_{B} c_{s}$. Here we assume at
$r=a$ the constant-charge boundary condition, and impose with the
second boundary condition at $r=\Rws$ the electroneutrality of the
cell. As input parameter we have $Z$, $\kres$, $a$, $\lambda_{B}$ and
$\Rws$. But inspection of eq.~(\ref{eq:1}) reveals that in fact only
three parameters are really independent, which are $\kres a$
(reservoir salt concentration), $Z \lambda_{B}/a$ (colloidal charge)
and $R/a$ (volume per colloid, i.e., volume fraction). Note that we
have also tacitly assumed the electrostatic potential to vanish in the
reservoir, where $n_{\pm}$ then becomes equal to $c_{s}$. The
microionic {\em charge} density $\rho(r)$ can be written as
\begin{equation}
  \label{eq:1a}
\rho(r) = n_{+}(r) - n_{-}(r) = - \frac{\kres^{2}}{4 \pi \lambda_{B}} \sinh \phi(r)  
\end{equation}
and the microionic {\em particle} density as
\begin{equation}
  \label{eq:1aa}
\rho_{T}(r) = n_{+}(r) + n_{-}(r) = 
\frac{\kres^{2}}{4 \pi \lambda_{B}} \cosh \phi(r) \:.
\end{equation}
Hence,
\begin{equation}
  \label{eq:1b}
  Z = 4 \pi \int_{a}^{R} dr \: r^{2}\:\rho(r)
\end{equation}
due to the electroneutrality of the cell, while the total number
$2N_{s} + Z$ of microions in the cell is obtained from
\begin{equation}
  \label{eq:1c}
2 N_{s} + Z = 4 \pi \int_{a}^{R} dr \: r^{2} \:\rho_{T}(r) 
\end{equation}
where $N_{s}$ is the number of pairs of salt ions in the cell.  We
define the (''nominal'') salt concentration in the system as $n_{s} =
N_{s}/V$ where $V = 4 \pi \Rws^{3}/ 3$ is the volume of the WS cell.
This volume should not be confused with that accessible to the
microions $V_{\free}=4 \pi (\Rws^{3}-a^{3})/ 3=V (1-\eta)$. The
''net'' salt ion concentration in the system then reads
$n_{s}/(1-\eta)$, and due to the Donnan effect we have $n_{s}/(1-\eta)
\le c_{s}$.

Here is the point where we can describe Alexander's {\it et al.}
recipe for charge
renormalization: 
\begin{itemize}
\item[i)] solve the full non-linear problem,
eq.~(\ref{eq:1}), find the potential at $\Rws$, $\phi_{\Rws}$, and the
total microion density, $n_{\pm}^{\Rws} = c_{s}e^{\mp\phi_{\Rws}}$,
\item[ii)] define the inverse screening constant from the microion density
at WS boundary,
\begin{equation}
  \label{eq:2}
\kpb^{2} = 4 \pi \lambda_{B}\Big( n_{+}^{\Rws} + n_{-}^{\Rws} \Big) 
= \kres^{2} \cosh\phi_{\Rws}\:,
\end{equation}
\item[iii)] linearize the boundary value problem in eq.~(\ref{eq:1}) about
$\phi=\phi_R$, determine the potential solution of the linearized PB
equation such that linear and non-linear solution match up to the
second derivative at $r=\Rws$,
\item[iv)] compute the effective charge
$Z_{\eff}$ from this solution by integrating the charge density
associated with the linear solution from $a$ to $R$. 
\end{itemize}
Replacing now
($Z$, $\kres$) by ($Z_{\eff}$, $\kpb$) in the effective Yukawa
pair-potential of the DLVO (Derjaguin-Landau-Verwey-Overbeek) theory
\cite{verwey},
\begin{equation}
  \label{eq:30}
  \beta v(r) = Z_{\eff}^{2} \lambda_{B} \left(\frac{e^{\kpb
  a}}{1+\kpb a}\right)^{2} \frac{e^{-\kpb r}}{r} \:,
\end{equation}
one has succeeded in retaining an effective pair-interaction of the
simple Yukawa form, even in cases where a high value of the bare
charge $Z$ may violate the condition for the linearization
approximation which this interaction potential is based on.
While the relevance of  defining an effective potential as 
(\ref{eq:30}) may be questioned at not too low density, 
the effective charge $Z_{\eff}$ and inverse screening length
$\kpb$ are defined without ambiguity. It is also noteworthy 
that as an exact property of the cell model under study \cite{Lin},
$\kpb$ is related to the pressure $P$ through
\be
\beta P = \frac{1}{4 \pi \lambda_B}\, \kpb^2.
\label{eq:P}
\ee

\section{Alexander and collaborators' prescription revisited}
\label{sec:alex}

In the original paper of Alexander et al \cite{alexander}, the
procedure just described was introduced as a numerical recipe, and it
was not pointed out that most of this scheme can actually be performed
analytically. We now describe the more direct way to calculate
Alexander's et al effective charge which needs as input nothing but
$\phi_{\Rws}$ from the solution of eq.~(\ref{eq:1}).

Let us consider the linearized version of eq.~(\ref{eq:1}).  We
linearize the PB equation about $\phi_{\Rws}$ and find
\begin{equation}
\label{eq:3}
\nabla^{2} \widetilde{\phi}(r)  =  \kres^{2} 
\left[\gamma_{0} +\widetilde{\phi}(r)\right]
\cosh \phi_{\Rws} 
\end{equation}
where $\widetilde{\phi} = \phi - \phi_{\Rws}$ and $\gamma_{0} = \tanh
\phi_{\Rws}$.  We observe that the inverse screening length in
eq.~(\ref{eq:3}) is given by $ \kres^{2} \cosh \phi_{\Rws} $ which is
just $\kpb^{2}$ from eq.~(\ref{eq:2}), and thus just the value
required in the Alexander et al.  prescription. This observation shows that
Alexander's prescription, in essence, amounts to a certain
linearization approximation where the potential value about which to
linearize, is just the potential at the cell boundary $\phi_{\Rws}$.
Linearizing eq.~(\ref{eq:1a}) and (\ref{eq:1aa}), the microionic
charge and particle densities now take the form
\begin{eqnarray}
  \label{eq:3a}
\rho(r) & = & 
- \frac{\kpb^{2}}{4 \pi \lambda_{B}} (\gamma_{0} +
\widetilde{\phi}(r))\\
  \label{eq:3b}
\rho_{T}(r) & = & 
\frac{\kpb^{2}}{4 \pi \lambda_{B}} (1 + \gamma_{0} \widetilde{\phi}(r))\:.
\end{eqnarray}
With eq.~(\ref{eq:2}), $\gamma_{0}$ can be rewritten as follows
\begin{equation}
  \label{eq:4}
\gamma_{0}  =  \tanh \phi_{\Rws} = \sqrt{1 -
  \left(\frac{\kres}{\kpb}\right)^{4}}  \:.
\end{equation}
Demanding that $\widetilde{\phi}(r)$ as well as its derivative are
zero at $r=\Rws$, we make sure that the solutions of the linearized
and fully non-linear PB equation agree at the cell edge
as required in Alexander's et al. scheme. Taken together the boundary value
problem of the linearized problem now reads,
\begin{equation}
\label{eq:5}
\begin{array}{rcl@{\hspace{1cm}}l}
\nabla^{2} \widetilde{\phi}(r) & = & \kpb^{2}  (\gamma_{0} +
 \widetilde{\phi}(r))  & a< r <\Rws \\
\widetilde{\phi}(r) & = & 0  & r = \Rws \\
\vec{n} \cdot \nabla \widetilde{\phi}(r) & = & 0 & r = \Rws
\end{array} 
\end{equation}
Note that eq.~(\ref{eq:1}) is a two-point boundary value problem,
while eq.~(\ref{eq:5}) represents a one-point boundary value problem.

\subsection{Effective colloidal charges}

The solution of eq.~(\ref{eq:5}) is given by
\begin{equation}
  \label{eq:6}
  \widetilde{\phi}(r) = \gamma_{0}\Big[
-1 + f_{+}\frac{e^{\kpb r}}{r}+ f_{-}\frac{e^{- \kpb r}}{r}
\Big]
\end{equation}
with 
\begin{equation}
f_{\pm} = \frac{\kpb \Rws \pm 1}{2 \kpb} \exp(\mp \kpb \Rws).
\end{equation}
With eq.~(\ref{eq:6}) we can now easily calculate the effective charge by
following eq.~(\ref{eq:1b}) and integrating eq.~(\ref{eq:3a}) over the
cell volume, resulting in the simple formula
\begin{equation}
  \label{eq:7}
Z_{\eff} = \frac{\gamma_{0}}{\kpb \lambda_{B}}
\Big\{ (\kpb^{2} a \Rws - 1)\sinh[\kpb(R-a)] + \kpb(R-a) \cosh[\kpb
(R-a)] \Big\}  \:.
\end{equation}
By comparison with the numerically calculated effective charges from
Alexander's et al. original prescription, we have explicitly checked that
this formula indeed produces Alexander's effective charges.  Note that
eq. (\ref{eq:7}) also follows from Gauss' theorem applied at the
surface of the colloid.

In summary, this then is the simple procedure to calculate
renormalized charges: 
\begin{itemize}
\item[i)] solve eq.~(\ref{eq:1}) to obtain
$\phi_{\Rws}$, 
\item[ii)] calculate $\kpb^{2} = \kres^{2} \cosh\phi_{\Rws}$,
\item[iii)] insert this into eq.~(\ref{eq:7}) to obtain $Z_{\eff}$.  
\end{itemize}
We emphasize that once eq.~(\ref{eq:1}) has been solved numerically, no
further numerical fitting procedure is required to match the
electrostatic potential from the linear and the non-linear solutions
at $r=\Rws$! In addition, the numerical solution of eq.~(\ref{eq:1})
is much simplified by taking account of the technical points described
in appendix \ref{app:a} (see also appendix \ref{app:b} for the salt free case). 

\begin{figure}
\includegraphics[width=\textwidth]{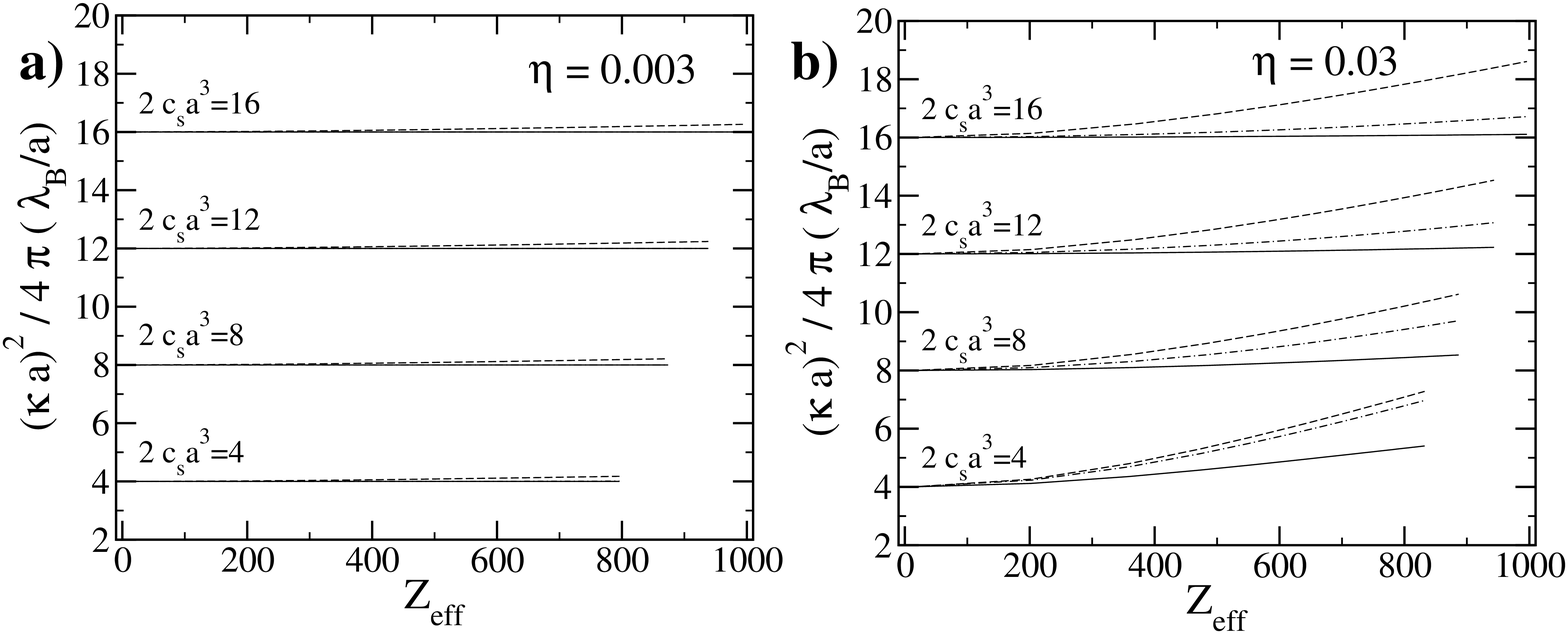}
\caption{\label{fig1}
  The inverse screening length $\kappa$ as a function of Alexander's
  et al.
  effective charge $Z_{\eff}$ [eq.~(\ref{eq:7})] for various reservoir
  salt concentrations $c_{s}$ and two different colloid volume
  fractions [$\eta = 0.003$ in a), and $\eta=0.03$ in b)]. Solid line:
  $\kpb^{2}$ from eq.~(\ref{eq:2}), dashed line: $\kappa^{2}$ from
  eq.~(\ref{eq:11}), dashed-dotted line: $\kpb^{2}|_{\gamma_{0}\to 0}$
  from eq.~(\ref{eq:10}). All curves terminate when the effective charge has
  reached its saturation value.}
\end{figure}

\subsection{Effective salt concentration}
\label{ssec:salt}

Alexander's original paper contains also a prescription how to
calculate the effective salt concentration $n_{s}^{\eff}$.  Again, we
here can derive a simple formula. Integrating the particle density in
eq.~(\ref{eq:3b}) over the cell volume accessible to the microions, as
in eq.~(\ref{eq:1c}), we obtain
\begin{equation}
  \label{eq:8}
2 N_{s}^{\eff} + Z_{\eff} =   
\frac{\kpb^{2}}{4 \pi \lambda_{B}} V_{\free}  
(1 -\gamma_{0}^{2}) + Z_{\eff} \gamma_{0}
\end{equation}
with $Z_{\eff}$ from eq.~(\ref{eq:7}), so that
\begin{equation}
  \label{eq:9}
2 n_{s}^{\eff} = \frac{\kpb^{2}}{4 \pi \lambda_{B}} (1
-\gamma_{0}^{2}) (1-\eta) - Z_{\eff} n_{c} (1-\gamma_{0}) \:,
\end{equation}
where $n_{c}=1/V$ is the colloid density in the suspension and
$n_{s}^{\eff}=N_{s}^{\eff}/V$.  This formula is again successfully
checked against the numerical values in Alexander's et al.  work. If
$\gamma_{0} \to 0$, eq.~(\ref{eq:9}) can be rearranged to give
\begin{equation}
  \label{eq:10}
\kpb^{2}\Big|_{\gamma_{0}\to 0}= \frac{1}{1-\eta} \,
4 \pi \lambda_{B} (Z_{\eff} n_{c} + 2 n_{s}^{\eff}) \:.
\end{equation}

Many studies, applying Alexander's et al. renormalization concept, can be
found in literature where $\kpb$ is computed not from
eq.~(\ref{eq:2}), as it should, but from
\begin{equation}
  \label{eq:11bis}
\kappa^{2} = \frac{1}{1-\eta} \,4 \pi \lambda_{B} (Z_{\eff} n_{c} + 2 n_{s}) \:,
\end{equation}
or
\begin{equation}
  \label{eq:11}
\kappa^{2} =  \,4 \pi \lambda_{B} (Z_{\eff} n_{c} + 2 n_{s}) \:,
\end{equation}
with $Z_{\eff}$ from Alexander's prescription and with $n_{s}$ being
the actual (as opposed to ''effective'') salt concentration in the
system. We here learn that this is certainly not the same as the
$\kpb$ from eq.~(\ref{eq:2})! Indeed, Eqs.~(\ref{eq:11bis}) and
(\ref{eq:11}) rely on two further assumptions, namely that
$\gamma_{0}$ can be set to zero and that $n_{s} \approx n_{s}^{\eff}$.
Fig.~(\ref{fig1}) checks how good these approximations are.  We
restrict the following discussion to low values of the volume fraction
$\eta$ so that expressions (\ref{eq:11bis}) and (\ref{eq:11}) coincide
(the rather formal case of larger $\eta$ will be addressed in section
\ref{ssec:nosalt}).  For two volume fractions and various reservoir
salt concentrations, we calculated in Fig.~(\ref{fig1}) the screening
factor as a function of Alexander's et al.  effective charge obtained from
eq.~(\ref{eq:7}), (i) using eq.~(\ref{eq:2}) (solid line), (ii) using
the formula $4 \pi \lambda_{B}(Z_{\eff} n_{c} + 2 n_{s}^{\eff})$ from
eq.~(\ref{eq:10}) with $n_{s}^{\eff}$ from eq.~(\ref{eq:9})
(dashed-dotted line) and (iii) using $4 \pi \lambda_{B}(Z_{\eff} n_{c}
+ 2 n_{s})$ from eq.~(\ref{eq:11}) with $n_{s}$ from the non-linear
calculation, i.e.  from eq.~(\ref{eq:1c}) (dashed line). As $Z_{\eff}$
goes to zero, $n_{s} \to n_{s}^{\eff}$, $n_{s} \to c_{s}$ and
$\phi_{\Rws} \to 0$, so all three expressions must lead to the same
$\kappa^{2} = 8 \pi \lambda_{B} c_{s} = \kres^{2}$. To be specific, we
took typical values of aqueous colloidal systems for $a$ and
$\lambda_{B}$ ($a=60$ nm, $\lambda_{B} = 0.713$ nm).  If however one
wants to be a more general, one has to specify just $\kres a$, $Z
\lambda_{B}/a$ and $\eta$ as these are the independent parameters of
the problem.  Multiplying the values of the x-axis with
$\lambda_{B}/a$ and those of the y-axis with $4 \pi \lambda_{B}/a$
where $\lambda_{B}/a =0.713/60=0.012$, one can transform
Fig.~(\ref{fig1}) into a plot that is valid for systems characterized by
other values of $a$ and $\lambda_{B}$.  The two colloid volume
fractions considered in Fig.~(\ref{fig1}) are both easily
experimentally realizable.  Note that the curves terminate at
different $Z_{\eff}^{\sat}$ which is due to the fact that the
saturation value of the effective charge depends on the salt content
of the suspension \cite{trizac2}.

It is evident from the figure that for low volume fraction ($\eta =
0.003$), it is of no consequence if the screening factor is calculated
from eq.~(\ref{eq:2}) or eq.~(\ref{eq:11}), and the error is certainly
negligible. At higher volume fraction, however, it is seen that both
approximations involved in taking eq.~(\ref{eq:11}) instead of
eq.~(\ref{eq:2}) -- namely, $n_{s}\approx n_{s}^{\eff}$, and
$\gamma_{0} \approx 0$ -- take effect: both formulae,
eq.~(\ref{eq:10}) and (\ref{eq:11}), fail to give the correct value
for $\kpb$ at low salt ($2c_{s}a^{3} = 4$) and high $Z_{\eff}$, but
the agreement between eq.~(\ref{eq:10}) (dashed-dotted line) and
$\kpb$ (solid line) improves if $c_{s}a^{3}$ increases, while
eq.~(\ref{eq:11}) still remains a rather poor approximation of $\kpb$.
This means that at high salt concentration $\gamma_{0} \approx 0$
produces only a small error, while $n_{s} \approx n_{s}^{\eff}$ is
always a bad approximation at high volume fraction, regardless the
value of $c_{s}a^{3}$. We will refer to expression 
(\ref{eq:11}) as a "naive" inverse screening length.
While such an estimation is inappropriate in the context of
Alexander and collaborators' scheme, it is noteworthy that it
naturally arises from a statistical mechanics treatment
of electrostatic interactions in colloidal suspensions
\cite{Chan1,Chan2}. This treatment is however performed
within a linear theory formalism and therefore discards the non linear effects
we are interested  in the present article.

Often, $n_{s}$ is known from the experiment, while $c_{s}$ (and thus
$\gamma_{0}$) is not, so that it seems to be difficult to calculate
$n_{s}^{\eff}$ from eq.~(\ref{eq:9}). However, it is not. One
can obtain $n_{s}$ as a function of $c_{s}$, from the solution of
eq.~(\ref{eq:1}) by means of eq.~(\ref{eq:1aa}) and (\ref{eq:1c}). In
cases where only $n_{s}$ is known from the experiment, one can then
use this curve to find the $c_{s}$ corresponding to the known $n_{s}$.
In fact, it is immaterial whether or not the experiment was actually
performed with the system coupled to a particle reservoir. This
becomes clear from the following consideration. Take a system coupled
through a semi-permeable membrane to a salt reservoir with salt
concentration $c_{s}$, and allow for some time till the Donnan
equilibrium is reached. Then, the salt concentration in the system is
$n_{s}/(1-\eta) \le c_{s}$. Now, replace the semi-permeable wall by a
unpenetrable wall and decouple the reservoir. The
microion-distribution between the colloids and thus the screening
factors will not change.  In other words, in cases where the system is
not coupled to a reservoir one can find a reservoir with an
appropriately chosen $c_{s}$ which when coupled to the system would
leave the microion density distribution in the system unaltered. This
means that all our considerations presented are also valid for
experiments in which the system is not coupled to a reservoir. This, 
of course, is strictly true only at a mean-field level of
description. Practically, one then has to proceed as follows: (i)
start from a trial value for $c_{s}$ and thus for $\kres^{2} = 8 \pi \lambda_{B}
c_{s}$, (ii) solve eq.~(\ref{eq:1}), (iii) calculate $n_{s}$ from
eq.~(\ref{eq:1aa}) and (\ref{eq:1c}), (iv) vary $c_{s}$ and repeat (i)
to (iii) until a $c_{s} = c_{s}^{exp}$ is found which leads to a
$n_{s}$ that equals the $n^{exp}_{s}$ from the experiment. The pair
$(n_{s}^{exp}, c_{s}^{exp})$ can now be used in all the formulae
presented above.

We close this section with a rather general remark. There is actually no need
to linearize about the potential at the WS cell edge as done in
eq.~(\ref{eq:5}). An alternative is suggested by the following
observation. Insert eq.~(\ref{eq:1aa}) into eq.~(\ref{eq:1c}) and
linearize about a potential $\bar{\phi}$,
\begin{eqnarray}
  \label{eq:12}
2 N_{s} + Z & = & \frac{\kres^{2}}{\lambda_{B}} 
\int_{a}^{R}dr\: r^{2} \: \cosh \phi(r)  \nonumber \\
& \approx & 
\frac{\kres^{2}}{\lambda_{B}} 
\int_{a}^{R}dr\: r^{2} \: \Big[
\cosh \bar{\phi} + \sinh \bar{\phi} (\phi - \bar{\phi}) \Big] \:.
\end{eqnarray}
If one now chooses
\begin{equation}
  \label{eq:12a}
\bar{\phi} = \frac{4 \pi}{V_{\free}} \int_{a}^{R} dr\: r^{2} \phi(r) \:,
\end{equation}
then
\begin{equation}
  \label{eq:13}
2 N_{s}^{\eff} + Z_{\eff}  = \frac{\kres^{2}}{4 \pi \lambda_{B}}
V_{\free} \cosh \bar{\phi}   
\end{equation}
so that with $\kappa^{2} = \kres^{2} \cosh \bar{\phi}$ as in
eq.~(\ref{eq:2}) one obtains
\begin{equation}
  \label{eq:14}
\kappa^{2} = \frac{1}{1-\eta}\,
4 \pi \lambda_{B} (Z_{\eff}n_{c} + 2 n_{s}^{\eff})  \:.
\end{equation}
In words: linearizing not about the cell edge value of the potential,
but about the average value of the potential in the cell
(eq.~(\ref{eq:12a})), leads to effective salt concentrations and
effective charges which -- when used to calculate the effective ionic
strength -- can be directly related to the inverse screening length in
the way given by eq.~(\ref{eq:14}), familiar from the Debye-H\"uckel
theory. This linearization scheme is worked out in
\cite{Trizacbis,deserno,Tamashiro}, but does not correspond to the original
proposal of Alexander {\it et al.}

\subsection{Situation without added salt}
\label{ssec:nosalt}
\begin{figure}
\includegraphics[width=10cm]{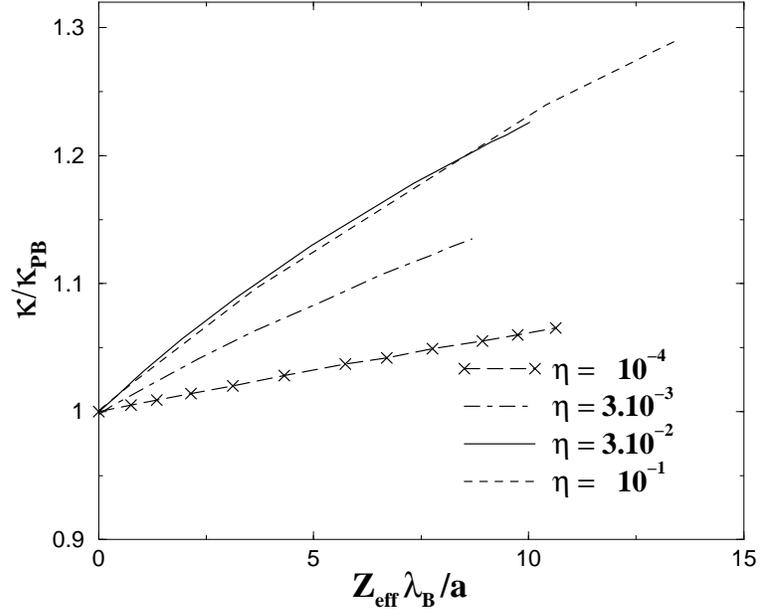}
\caption{\label{fig2}
  Ratio of the ``naive'' inverse screening length $\kappa$ as given in
  eq.~(\ref{eq:kapp}) over $\kpb$ defined from the microion density at
  the WS cell boundary, eq.~(\ref{eq:26}), as a function of
  Alexander's 
  et al. rescaled
  effective charge $Z_{\eff} \lambda_B/a$ [eq.~(\ref{eq:7})], for
  various colloid packing fractions $\eta$ in the salt free case.  The
  curves terminate at different $Z_{\eff}^{\sat} \lambda_B/a$ since
  the saturation value of the effective charge depends on the packing
  fraction \cite{Donnan}.}
\end{figure}

In the limit where the system is in osmotic equilibrium with a
reservoir of vanishing salt density ($c_s \to 0$), or in the canonical
situation where no salt is added to the solution, the PB equation takes
the form
\be
\nabla^{2} \phi(r)  =  -\mu^2 e^{-\phi(r)},
\label{eq:25}
\ee
where $\mu$ is a prefactor whose value is determined through the
electroneutrality constraint. We have
\be
\kpb^2 = 4 \pi \lambda_{B} n_{+}^{\Rws} \, = \,\mu^2 \,e^{-\phi_{\Rws}}.
\label{eq:26}
\ee 
Equation (\ref{eq:7}), which provides the connection between the
relevant screening parameter $\kpb$ and the effective charge
$Z_{\eff}$, is still correct if $\gamma_0=1$ \cite{trizac2}.
It is then tempting to compare $\kpb$ to the counterpart of expression
(\ref{eq:10}):
\be \kappa^2 = \frac{1}{1-\eta} \, 4 \pi \lambda_{B}
Z_{\eff} n_c.
\label{eq:kapp}
\ee
This comparison is shown in Fig. \ref{fig2}. Over a wide range of
packing fractions, the ratio $\kappa/\kpb$ deviates from unity, but
less than 30\%. At relatively large bare charges, $Z$ and $Z_{\eff}$
differ significantly, so that neglect of charge renormalization in
computing the screening parameter $\kappa$ [substitution of $Z_{\eff}$
by $Z$ in eq. (\ref{eq:kapp})] badly fails compared to the ``exact''
$\kpb$.

It is instructive to investigate analytically a few limiting cases.
We have computed $\kpb R$ from the solution of the non-linear PB. At fixed
volume fraction, this quantity is bounded from above by its saturation
value obtained when $Z\to\infty$.  In the limit where the packing
fraction $\eta$ vanishes, this saturation value $\kpb^{\sat} R$ is
observed to vanish (although very slowly, as $-\eta^{1/3} \log\eta)$.
Hence, $\kpb a = \eta^{1/3} \kpb \Rws$ also vanishes, and this piece
of information allows to linearize the exact relation (\ref{eq:7}),
which then takes a simple form:

\be
Z_{\eff} \frac{\lambda_B}{a} \,\sim\, \frac{1}{3} \, (\kpb \Rws)^2 \, \eta^{-1/3},
\ee
where the symbol $\sim$ stands for ``asymptotically equivalent''.
In other words, we have
\be
\kpb^2 \,\, \stackrel{\eta \to 0}{\sim} \,\, 4 \pi \lambda_{B} Z_{\eff} n_c,
\label{eq:28}
\ee
so that $\kappa$ from eq. (\ref{eq:kapp}) and $\kpb$ coincide in this
limit.  It may be observed in Fig. \ref{fig2} that $\eta=10^{-4}$ is
not small enough to get the limiting behavior $\kappa/\kpb=1$,
irrespective of $Z$ (and thus $Z_{\eff}$).  In the opposite (and
academic) limit where $\eta \to 1$, ({\it i.e.} $\Rws \to a$), we
obtain from (\ref{eq:7})
\be
\frac{\kappa}{\kpb}  \,\, \stackrel{\eta \to 1}{\to} \,\, 1.
\ee
This last result somehow illustrates the relevance of the factor
$1-\eta$ in (\ref{eq:kapp}). If the microions densities would have been
defined with respect to the total volume of the cell and not the
sub-volume accessible to microions, we would have obtained the
alternative naive expression for the screening parameter
\be
(\kappa')^2 = \, 4 \pi \lambda_{B} Z_{\eff} n_c,
\ee
so that 
\be
\frac{\kappa'}{\kpb}  \,\, \stackrel{\eta \to 1}{\sim} \,\, 
\sqrt{3(\eta^{2/3}-\eta)} \to 0.
\ee
Alternatively, in the limit of low bare charge $Z$ we get
\be
Z_{\eff} \frac{\lambda_B}{a}\,\,  \stackrel{Z \to 0}{\sim} \,\,  
\frac{1}{3} \, 
\left(\frac{1}{\eta}-1\right)\, (\kpb\, a)^2.
\ee
As a consequence, 
\be
\frac{\kappa}{\kpb}  \,\,  \stackrel{Z \to 0}{\to} \,\, 
1.
\ee
This feature may be observed in Fig \ref{fig2} and is compatible with 
the result of eq. (\ref{eq:28}). 

With added salt, similar argument may be put forward to provide an
analytical relation between $\kpb/\kres$ and the density for e.g.
large dilutions. Such expressions are nevertheless physically less
transparent than those derived here for de-ionized suspensions, and
have been omitted.

\section{Conclusion}

For colloidal spheres, we have reconsidered the original charge
renormalization prescription proposed by Alexander {\it et al.}
\cite{alexander}. The computation of renormalized charge and salt
content has been simplified in two respects: {\em a)} by the
derivation of analytical expressions giving effective quantities as a
function of parameters that are directly obtained from the solution of
the non-linear PB problem; {\em b)} by converting the initial
two-point boundary value non-linear PB problem into a computationally
more convenient one-point boundary value problem (see appendices
\ref{app:a} and \ref{app:b}).  While we have restricted here to
spherically symmetric polyions, similar considerations may be applied
to the cylindrical geometry, relevant, e.g., to understand properties
of solutions of charged polymers as for example the DNA molecule.

Once the effective charge and salt density are known, a ``naive''
screening parameter $\kappa$ may be defined as $\kappa^2 = 4 \pi
\lambda_B (Z_{\eff} n_c + 2 n_s^{\eff})/(1-\eta)$, where the factor
$1-\eta$ accounts for the fact that a fraction $\eta$ of the
Wigner-Seitz cell is not accessible to the microions.  Strictly
speaking, this inverse Debye length does not coincide in general
with the relevant screening parameter $\kpb$, which has to be defined
from the microions density at the WS boundary. In all the
cases investigated here, the difference between $\kappa$ and $\kpb$
was less than 40\%. Moreover, in the salt free case, $\kappa$ and
$\kpb$ have been shown analytically to coincide for both low and large
packing fractions (irrespective of the charge), and also for vanishing
bare charges.  This implies that the ``naive''expression $\beta P =
(Z_{\eff} n_c + 2 n_s^{\eff})/(1-\eta)$ yields a reasonable zeroth
order equation of state for the suspension.  On the other hand,
neglect of charge renormalization, which amounts to defining $\kappa$
or $\beta P$ through the bare charge and salt density, appears to
provide an extremely poor approximation for both $\kpb$ and the
pressure.

\medskip

Acknowledgments: We would like to thank Yan Levin and Jure Dobnikar
for interesting discussions, and Mario Tamashiro for a careful reading
of the manuscript.

\begin{appendix}
\section{Numerical procedure with added electrolyte}
\label{app:a}
In this appendix, we propose a few
Mathematica$^{\hbox{\scriptsize \textregistered}}$ lines of code to solve
PB equation for a charged sphere in a concentric spherical Wigner-Seitz 
cell. To this end, all distances are rescaled with the diameter $a$
of the colloid, and the charge expressed in units of $a/\lambda_B$. 
It is convenient to recast the initial two point boundary value 
problem (\ref{eq:1}) into a one point boundary value problem
by assigning an {\it a priori} value $\phi_{test}$ to the rescaled potential
at WS boundary $\phi_R$. For the situation where $\kres a = 2.0$ at a volume
fraction $\eta=0.1$, the following procedure finds the corresponding
solution (with the arbitrary choice $\phi_{test}=0.1$):
\begin{eqnarray}
&&\kres =      2.0  \   ; \quad  \eta = 0.1  \  ; \quad
R = \eta^{-1/3} \ ;\quad \phi_{test} = 0.1 \ ;\quad sol =  \label{eq:math}\\
&&
  {\tt NDSolve}[\,\{\,  2 \phi'[r] + r \phi''[r] == \kres^2 r\, 
  {\tt Sinh}[\phi[r]], \phi[R] == \phi_{test},  \phi'[R] == 0 \,\},  \nonumber \\
&& \qquad \qquad \qquad  \phi, \{r, 1, R\}, {\tt WorkingPrecision} \to 20\,]; 
\qquad\qquad  \nonumber
\end{eqnarray}
The potential $\phi$ may then be visualized as a function of $r/a$ 
with the command
\be
{\tt Plot}[\, {\tt Evaluate}[ \phi[r] /. sol ], \{r, 1, R\} \,]
\ee
and the bare charge corresponding to the specific choice made for $\phi_{test}$
is obtained as the result of
\be
(\,{\tt Evaluate}[\, \phi'[1] /. sol \,]\,)\,[[1]].
\ee     
With the above parameters, we get 2.55, which corresponds to the value
of $Z \lambda_B/a$. 

\begin{figure}
\includegraphics[width=10cm]{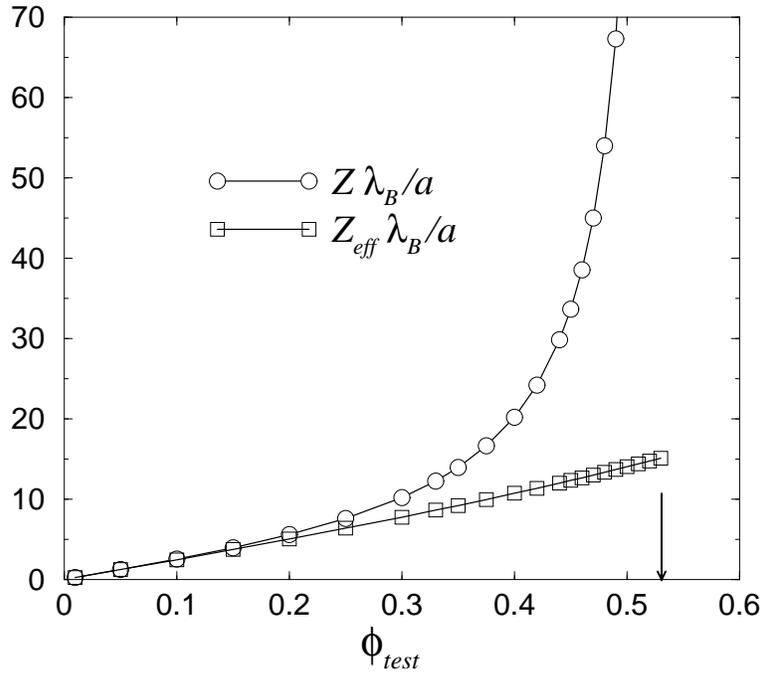}
\caption{\label{fig:10}
  Bare and effective charges as a function of the ``guess'' boundary potential
  $\phi_{test}$. Here, $\kres a = 2.0$ and the volume fraction is $10\%$.
  The arrow indicates the saturation value for $\phi_{test}$. }
\end{figure}

The limit $\phi_{test} \to 0$ corresponds to the 
limit $Z\to 0$. However, the solution of our one point boundary value
problem only exists for $\phi_{test} \leq \phi_{sat}$, and the limit
$\phi_{test} \to \phi_{sat}$ corresponds to $Z \to \infty$ which is equivalent to
$Z_{\eff} \to Z_{\eff}^{\sat}$. Consequently, starting from low values
of $\phi_{test}$, the solution associated with a targeted $Z$ is 
easily found by dichotomy, adjusting iteratively the values of $\phi_{test}$. 
For any value of $\phi_{test}$, $\kpb$ follows from eq. (\ref{eq:2}) and 
$Z_{\eff}$ is computed invoking eq. (\ref{eq:7}). For the parameters
used in the example (\ref{eq:math}), the relation between $Z$, $Z_{\eff}$
and $\phi_{test}$ is illustrated in Figure \ref{fig:10}. 

\section{Numerical procedure without added electrolyte}
\label{app:b}
In the salt-free case, it is also convenient to rephrase the problem
under study as a one point boundary value problem.  At fixed bare charge,
a possibility would be to consider the limit $\kres \to 0$ of a system
in contact with a salt reservoir. In this (formally correct) limit,
the corresponding $\phi_{test}$ diverges, and it turns out to be
more convenient to impose that $\phi(R)=0$.  This choice is such that
$\mu = \kpb$ with the notations of eq.  (\ref{eq:26}). Poisson's
equation (\ref{eq:25}) is then solved once a trial value has been
chosen for $\mu$. For instance, with $\eta=0.1$ and $\mu = 1.1$, the
Mathematica$^{\hbox{\scriptsize \textregistered}}$ line
\begin{eqnarray}
&&\mu =  1.1 \ ; \quad \eta = 0.1   \ ; \quad   R =  \eta^{-1/3}  \   ; 
\label{eq:b1} \\
&&sol = {\tt NDSolve}[\,\{  2 u'[r] + r u''[r] 
             == \mu^2 r \,{\tt Exp}[u[r]], u[R] == 0, 
        u'[R] == 0\},\nonumber \\
&& u, \{r, 1, R\}, {\tt WorkingPrecision} \to 20\,]; \,\,
\phi[r\!\_] := ({\tt Evaluate}[ u[r] /. sol])[[1]]; \nonumber
\end{eqnarray}
allows to find the reduced potential $\phi$. The associated bare charge 
$Z\lambda_B/a$
is computed {\it a posteriori} from Gauss' theorem
\be
-({\tt Evaluate}[ \phi'[1]])
\ee
With the example described in eq. (\ref{eq:b1}), we get 5.217. 
Here, $\mu$ plays a similar role in the resolution as $\phi_{test}$
in appendix \ref{app:a}. The solution of eq. (\ref{eq:25}) only
exists for $\mu < \mu^{\sat}$, and when $\mu \to \mu^{\sat}$
({\it i.e.} when $\kpb \to\kpb^{\sat}$), the bare charge diverges. 
Implementing a PB-like cell problem along the lines described here
constitutes a substantial simplification with respect to the 
traditional route, usually involving a Fortran or {\tt C} code
with a numerical fitting procedure of the linear and non-linear 
solutions of PB theory.

\end{appendix}


\begin{thebibliography}{10}

\bibitem{reviewBelloni97}
L. Belloni, Colloids Surfaces A {\bf 140},  227  (1998).

\bibitem{levin:rev}
Y. Levin, Rep. Prog. Phys. {\bf 65},  1577  (2002).

\bibitem{Likos}
C.N. Likos, Phys. Rep.-Rev. Sec. Phys. Lett. {\bf 348},  267  (2001).

\bibitem{Quesada}
M. Quesada-Perez, J. Callejas-Fernandez, and R. Hidalgo-Alvarez, Adv. Colloid
  Interface Sci. {\bf 95},  295  (2002).

\bibitem{alexander}
S. Alexander, P.M. Chaikin, P. Grant, G.J. Morales, P. Pincus, and D. Hone, J.
  Chem. Phys. {\bf 80},  5776  (1984).

\bibitem{trizac1}
E. Trizac,   L. Bocquet and   M. Aubouy, 
Phys. Rev. Lett. {\bf 89}, 248301 (2002).   

\bibitem{trizac2}
L. Bocquet, E. Trizac and M. Aubouy, 
J. Chem. Phys. {\bf 117}, 8138   (2002).

\bibitem{JPA}
M. Aubouy, E. Trizac, L. Bocquet, to appear in J. Phys. A. (2003).


\bibitem{Behrens}
S.H. Behrens and D.G. Grier, J. Chem. Phys. {\bf 115},  6716  (2001).

\bibitem{Wette}
P. Wette, H.J. Sch\"ope, and T. Palberg, J. Chem. Phys. {\bf 116},  10981
  (2002).

\bibitem{vonGru3}
H.H. von Gr\"unberg, L. Helden, P. Leiderer, and C. Bechinger, J. Chem. Phys.
  {\bf 114},  10094  (2001).

\bibitem{Garbow}
N. Garbow, M. Evers, and T. Palberg, Colloid Surf. A-Physicochem. Eng. Asp.
  {\bf 195},  227  (2001).

\bibitem{Wette2}
P. Wette, H.J. Sch\"ope, R. Biehl, and T. Palberg, J. Chem. Phys. {\bf 114},
  7556  (2001).

\bibitem{Fernand}
A. Fernandez-Nieves, A. Fernandez-Barbero, and F.J. de~las Nieves, Phys. Rev. E
  {\bf 64},  032401  (2001).

\bibitem{Groenew}
J. Groenewold and W.K. Kegel, J. Phys. Chem. B {\bf 105},  11702  (2001).

\bibitem{Lobaski}
V. Lobaskin, A. Lyubartsev, and P. Linse, Phys. Rev. E {\bf 63},  020401
  (2001).

\bibitem{Terao2}
T. Terao and T. Nakayama, Colloid Surf. A-Physicochem. Eng. Asp. {\bf 182},
  299  (2001).

\bibitem{Anta}
J.A. Anta and S. Lago, J. Chem. Phys. {\bf 116},  10514  (2002).

\bibitem{Mukherj}
A.K. Mukherjee, K.S. Schmitz, and L.B. Bhuiyan, Langmuir {\bf 18},  4210
  (2002).

\bibitem{Ulander}
J. Ulander, H. Greberg, and R. Kjellander, J. Chem. Phys. {\bf 115},  7144
  (2001).

\bibitem{Allahya}
E. Allahyarov and H. L\"owen, J. Phys.: Condens. Matter {\bf 13},  L277  (2001).

\bibitem{Allahya2}
E. Allahyarov and H. L\"owen, Phys. Rev. E {\bf 63},  041403  (2001).

\bibitem{Burak}
Y. Burak and D. Andelman, J. Chem. Phys. {\bf 114},  3271  (2001).

\bibitem{Lukatsk}
D.B. Lukatsky and S.A. Safran, Phys. Rev. E {\bf 63},  011405  (2001).

\bibitem{Diehl}
A. Diehl, M.C. Barbosa, and Y. Levin, Europhys. Lett. {\bf 53},  86  (2001).

\bibitem{Schmitz1}
K.S. Schmitz, Phys. Rev. E {\bf 65},  061402  (2002).

\bibitem{vonGru}
H.H. von Gr\"unberg, R. van Roij, and G. Klein, Europhys. Lett. {\bf 55},  580
  (2001).
  
\bibitem{Tamashiro}
M.N. Tamashiro and H. Schiessel, e-print cond-mat/0210245.

\bibitem{Tellez}
G. T\'ellez and E. Trizac, J. Chem. Phys. at press, e-print cond-mat/0209114.

\bibitem{deserno}
M. Deserno and H.H. von Gr\"unberg, Phys. Rev. E {\bf 66},  011401  (2002).

\bibitem{Schmitz}
K.S. Schmitz, {\em Macroions in Solution and Colloidal Suspension} (VCH, New
  York, 1993).

\bibitem{Liu}
J.N. Liu, H.J. Sch\"ope, and T. Palberg, J. Chem. Phys. {\bf 116},  5901
  (2002).

\bibitem{Okubo}
T. Okubo, H. Kimura, T. Hatta, and T. Kawai, Phys. Chem. Chem. Phys. {\bf 4},
  2260  (2002).

\bibitem{Pertsin}
A. Pertsinidis and X.S. Ling, Phys. Rev. Lett. {\bf 8709},  098303  (2001).

\bibitem{Beching2}
C. Bechinger, M. Brunner, and P. Leiderer, Phys. Rev. Lett. {\bf 86},  930
  (2001).

\bibitem{Beching}
C. Bechinger and E. Frey, J. Phys.: Condens. Matter {\bf 13},  R321  (2001).

\bibitem{Nagele1}
G. N\"agele, M. Kollmann, R. Pesche, and A.J. Banchio, Mol. Phys. {\bf 100},
  2921  (2002).

\bibitem{Nguyen3}
T.T. Nguyen and B.I. Shklovskii, J. Chem. Phys. {\bf 114},  5905  (2001).

\bibitem{Nguyen}
T.T. Nguyen and B.I. Shklovskii, J. Chem. Phys. {\bf 115},  7298  (2001).

\bibitem{Nguyen2}
T.T. Nguyen and B.I. Shklovskii, Physica A {\bf 293},  324  (2001).

\bibitem{Schiess}
H. Schiessel, R.F. Bruinsma, and W.M. Gelbart, J. Chem. Phys. {\bf 115},  7245
  (2001).

\bibitem{Zhulina}
E. Zhulina, A.V. Dobrynin, and M. Rubinstein, Eur. Phys. J. E {\bf 5},  41
  (2001).

\bibitem{Levin}
Y. Levin, A. Diehl, A. Fernandez-Nieves, and A. Fernandez-Barbero, Phys. Rev. E
  {\bf 65},  036143  (2002).

\bibitem{DDeserno}
M. Deserno, Eur. Phys. J. E {\bf 6},  163  (2001).

\bibitem{Evilevi}
A. Evilevitch, V. Lobaskin, U. Olsson, P. Linse, and P. Schurtenberger,
  Langmuir {\bf 17},  1043  (2001).

\bibitem{Piazza}
R. Piazza and A. Guarino, Phys. Rev. Lett. {\bf 88},  208302  (2002).

\bibitem{Zagorod}
A.G. Zagorodny, A.G. Sitenko, O.V. Bystrenko, P.P.J.M. Schram, and S.A.
  Trigger, Phys. Plasmas {\bf 8},  1893  (2001).

\bibitem{Stern}
O. Stern, Ztschr. Elektrochem. {\bf 30}, 508 (1924).

\bibitem{Kalinin}
see e.g. V.V. Kalinin and C.J. Radke, Colloids Surf. A {\bf 114}, 337
(1996); P.A. Kralchevsky, K.D. Danov, G. Broze and A. Mehreteab, 
Langmuir {\bf 15},
2351 (1999).


\bibitem{trizac40}
J.P. Hansen and E. Trizac, Physica A {\bf 235},  257  (1997).

\bibitem{Dubois}  M. Dubois, T. Zemb, L. Belloni, A. Delville, P. Levitz and
R. Setton, J. Chem. Phys. {\bf 75}, 944 (1992).

\bibitem{Marcus}  R.A. Marcus,
J. Chem. Phys. {\bf 23}, 1057 (1955).  

\bibitem{DeHo01}
  M.\ Deserno and C.\ Holm, in: Proceedings of the NATO
  Advanced Study Institute on Electrostatic Effects in Soft Matter and
  Biophysics, ed.\ by C.\ Holm et al., Kluwer (2001).

\bibitem{verwey}
E.J.W. Verwey and J.T.G. Overbeek, {\em Theory of the Stability of Lyophobic
  Colloids} (Elsevier, Amsterdam, 1948).
  
\bibitem{Lin}
H. Wennerstr\"om, B. J\"onsson and P. Linse, 
J. Chem. Phys. {\bf 76}, 4665 (1982).   
  
\bibitem{Trizacbis}
E. Trizac and J.-P. Hansen,
Phys. Rev. E {\bf 56}, 3137 (1997).

\bibitem{Donnan}
E. Trizac, M. Aubouy and L. Bocquet, to appear in 
J. Phys.: Condens. Matter
 {\bf 15}, S291 (2003).

\bibitem{Chan1}
B. Beresford-Smith and D.Y.C. Chan, Chem. Phys. Lett. {\bf 92}, 474
(1982).

\bibitem{Chan2}
B. Beresford-Smith and D.Y.C. Chan, J. Coll. Int. Sci. {\bf 105}, 216
(1985).

\end{thebibliography}

\end{document}